\documentclass[twocolumn,english,prl,aps,showpacs,floatfix]{revtex4-1}
\usepackage[T1]{fontenc}
\usepackage[latin9]{inputenc}
\usepackage{verbatim}
\usepackage{amsmath}
\usepackage{amssymb}
\usepackage{graphicx}

\makeatletter

\providecommand{\tabularnewline}{\\}

\usepackage{babel}

\makeatother

\usepackage{babel}
\begin{document}

\title{On the Conservation of Information in Quantum Physics}

\author{Marco Roncaglia}
\email{marco.roncaglia.it@gmail.com}

\affiliation{Physics Department and Research Center OPTIMAS, University of Kaiserslautern,
Germany}

\date{\today}
\begin{abstract}
According to quantum mechanics, the informational content of isolated
systems does not change in time. However, subadditivity of entropy
seems to describe an excess of information when we look at single
parts of a composite systems and their correlations. Moreover, the
balance between the entropic contributions coming from the various
parts is not conserved under unitary transformations. Reasoning on
the basic concept of quantum mechanics, we find that in such a picture
an important term has been overlooked: the intrinsic quantum information
encoded in the coherence of pure states. To fill this gap we are led
to define a quantity, that we call coherent entropy, which is necessary
to account for the ``missing'' information and for re-establishing
its conservation. Interestingly, the coherent entropy is found to
be equal to the information conveyed in the future by quantum states.
The perspective outlined in this paper may be of some inspiration
in several fields, from foundations of quantum mechanics to black-hole
physics. 
\end{abstract}

\maketitle

\paragraph*{Introduction}

Every physicist is confident with the principle of energy conservation
and aware on its importance and implications. During time evolution
in isolated systems, energy is converted from one form to another
or transferred between different subsystems, provided the total amount
remains the same. However, when we consider information the picture
is not so clear. In quantum physics, the conservation of information
has been related to no-cloning theorems \cite{Horodecki05}, but apparently
it has not been associated to a suitable conserved quantity. It is
a well-known fact that the von-Neumann entropy $S(\rho)=-\mathrm{Tr}(\rho\log_{2}\rho)$
of any isolated quantum system with density operator $\rho$ does
not change in time. This is a consequence of the fact that $S(\rho)$
depends only on the spectrum of $\rho$, and the unitarity of time
evolution preserves the spectrum at the quantum level. In this sense,
people say that any physical process governed by quantum mechanics
information is never lost. However, this is a static vision that involves
isolated quantum systems. So far, there is no complete theory able
to clearly describe how quantum information ``flows'' between interacting
systems, accounting for a correct balance at any time. 

Whenever a system $A$, which is initially in a pure state $\rho_{A}=|\psi_{A}\rangle\langle\psi_{A}|$,
is no more isolated because it interacts with another system $B$,
we start to observe the increase of its mixedness, quantified by the
entropy $S(\rho_{A})$, where $\rho_{A}=\mathrm{Tr}_{B}(\rho_{AB})$.
If we look at the total entropy $S(\rho_{A})+S(\rho_{B})$, we see
that it can only increase with time as a consequence of the subadditivity
property $S(\rho_{AB})\leq S(\rho_{A})+S(\rho_{B})$ \cite{NielsenChuang_book,Preskill_notes}.
Moreover, interaction creates some correlations between $A$ and $B$,
with an additional contribution of information, which can be measured
by their mutual information $I_{A:B}$. Such a description is not
governed by a balance equation of a conserved quantity, giving the
illusion that information is created \emph{via} interaction. Where
does all this information come from? It is clear that entropy is not
the right quantity to describe the full informational content of quantum
states. 

In this paper, we propose to treat separately the coherent and the
incoherent contributions of the informational content of quantum states.
Starting from the very basic principles of quantum mechanics, we will
introduce the concept of coherent entropy, a quantity able to detect
the information that quantum states convey in time. In this context,
pure states contain more coherent information than mixed states, as
the missing information has been converted into correlations with
the environment. We will find that the coherent information, associated
to genuinely quantum phenomena, is indeed conserved under unitary
processes.

\paragraph{Informational content of quantum states }

A pure state is the eigenstate of some complete observable, whose
measurement gives a fully predictable outcome. Hence, the associated
zero entropy accounts for the absence of information obtainable by
a repeated measurement over many copies. However, it appears reductive
to attribute a zero informational content to pure states. In fact,
in contrast with single deterministic classical states, they represent
the ideal resource for performing quantum tasks, like interference
phenomena, quantum computation, and so on. An indicator beyond the
(von-Neumann) entropy is needed to describe such information. 

Let us take the textbook example of one qubit, i.e. a pure state in
dimension $d=2$ of the Hilbert space. Observables are represented
by the set of Pauli matrices with eigenvalues $\{+1,-1\}$, where
$\sigma^{z}$ has eigenstates $\{\left|\uparrow\right\rangle ,\left|\downarrow\right\rangle \}$,
and $\sigma^{x}$ has eigenstates $\{|+\rangle,|-\rangle\}$. If our
source emits quantum objects in the state $\left|\uparrow\right\rangle =(|+\rangle+|-\rangle)/\sqrt{2}$,
a measurement along $\sigma^{x}$ gives a random sequence of values
$+1$ and $-1$, with equal probability $p_{+}=p_{-}=1/2$. In this
case the entropy of information is one bit, i.e. the maximum obtainable
for a dichotomic variable. Differently, if the observer measures along
$\sigma^{z}$, he obtains the constant sequence of +1, with zero entropy.
This property of detecting different entropies under different measurements
is genuinely quantum, as it is ultimately due to interference: in
the present example, distinct states ($|+\rangle,|-\rangle$) coherently
recombine into a single state ($\left|\uparrow\right\rangle $), thanks
to their well-defined relative phases. 

\begin{figure}
\includegraphics[scale=0.5]{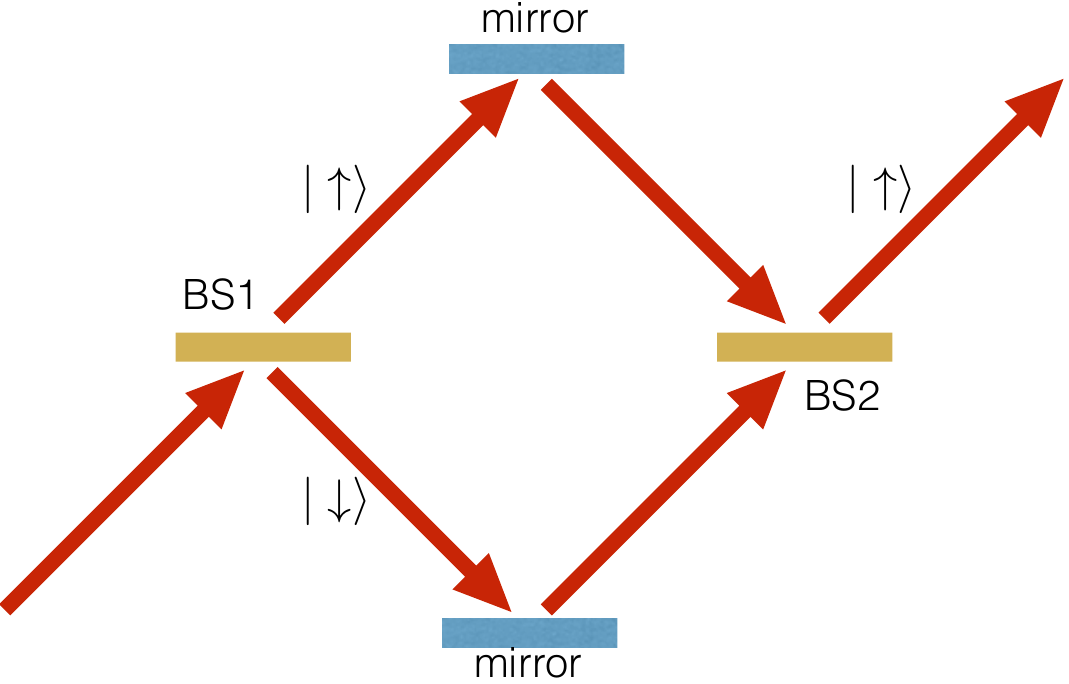}\caption{In a Mach-Zehnder interferometer single photons after the beam splitter
BS1 produce a signal with maximum entropy $S=1$. In fact, a photon
detection just after BS1 gives random sequences of $+1$ and $-1$
(photons in the upper and the lower branch, respectively), with equal
probability. Instead after BS2, the photons are all found in the $|\uparrow\rangle$
state, which yield a signal with zero entropy. Such a recombination
into a single state is a genuinely quantum phenomenon, since it is
due to interference between the two coherent beams coming from BS1.
\label{fig:Mach-Zehnder}}
\end{figure}

In quantum optics, this example is realized by the Mach-Zehnder interferometer
(see Fig.\ref{fig:Mach-Zehnder}), where the path of a single-photon
beam is split in two different directions by a 50\% reflective mirror
(the beam splitter BS1) and then constructively recombined into a
single path by a second beam splitter BS2. A measure of the presence
of the photon in the upper ($\left|\uparrow\right\rangle $) or lower
branch ($\left|\downarrow\right\rangle $) after BS1 gives a random
sequence of $+1$ and $-1$, with equal probability. At the output
after BS2, the photon detection produces a steady sequence of $+1$'s,
which yield a signal with zero entropy. As the role of the beam splitter
is to rotate the basis of measurement, we deduce that the entropy
of the detected signal depends in essence on the observable we choose. 

In the case of mixed states the effect of interference is reduced,
so in every measurement basis we expect to have a residual randomness
with a nonvanishing entropy; the limit case is the completely mixed
state, where the entropy of the outcome is maximal for every measurement.
Notice that an observer who detects the signal after BS1 is not able
to distinguish between the pure case and the totally mixed one as
they have the same statistics. However, it is important to introduce
a measure able to account for the information carried the in the former
coherent case, different from the entropy $S(\rho)$ which only quantifies
the incoherent information in the latter case. 

\paragraph{Coherent entropy}

Once the measurement basis is fixed, the probability of obtaining
a given output is encoded in the diagonal elements $\rho_{ii}$, $i=1,\dots,d$
of the density operator in that basis. The entropy of the output measurements
is given by the diagonal entropy of $\rho$, which we define as $S(\tilde{\rho})$,
where $\tilde{\rho}_{ij}=\delta_{ij}\rho_{ij}$ is the density operator
where all the off-diagonal entries have been set to zero. Now, we
define the \emph{coherent entropy} as 
\begin{equation}
S_{c}(\rho)=\max_{\sigma\in\mathcal{U}_{\rho}}[S(\tilde{\sigma})]-\min_{\sigma\in\mathcal{U}_{\rho}}[S(\tilde{\sigma})],\label{eq:Sc def}
\end{equation}
where $\mathcal{U}_{\rho}=\{U\rho U^{\dagger}:U\in\mathcal{M}_{d\times d},UU^{\dagger}=\mathbb{I}\}$
is the set of all matrices which are unitarily equivalent to $\rho$.
In other words, $S_{c}(\rho)$ measures the difference between the
maximal and the minimal entropy of the outputs obtained by measuring
$\rho$ over any possible observable. As we have seen, this difference
accounts for all the interference effects, so it has to be intended
as a measure \cite{Baumgratz14} of the coherent informational content
of $\rho.$ As it should be for a proper intrinsic property of a quantum
state, $S_{c}(\rho)$ is independent of any choice of measurement
made by the experimenter, i.e. it is invariant under local unitary
transformations. The apparently hard optimization problem of evaluating
Eq.(\ref{eq:Sc def}) eventually leads to a very simple result: 
\begin{equation}
S_{c}(\rho)=\log_{2}d-S(\rho).\label{eq:Sc}
\end{equation}
\emph{Proof} - For every density operator $\sigma\in\mathcal{U}_{\rho}$,
we have $S(\tilde{\sigma})=-\mathrm{Tr}(\tilde{\sigma}\log\tilde{\sigma})=-\mathrm{Tr}(\sigma\log\tilde{\sigma})$,
as $\tilde{\sigma}$ is the diagonal part of $\sigma$. The difference
\[
S(\tilde{\sigma})-S(\sigma)=\mathrm{Tr}[\sigma(\log\sigma-\log\tilde{\sigma})]=S(\tilde{\sigma}\parallel\sigma)\geq0
\]
due to the non negativity of the relative entropy $S(\tilde{\sigma}\parallel\sigma)$,
and the equality holds if $\tilde{\sigma}=\sigma$ \cite{Wehrl78}.
Since $S(\sigma)=S(\rho)$, $\forall\sigma\in\mathcal{U}_{\rho}$,
we get $\min_{\sigma\in\mathcal{U}_{\rho}}[S(\tilde{\sigma})]=S(\rho)$.
Intuitively, the operation of deleting the off-diagonal elements is
a decoherence operation (i.e. entropy increasing), which has no effect
only in the basis where $\rho$ is already diagonal. Regarding the
first term in Eq.(\ref{eq:Sc def}), we can say it is equal to the
entropy of the totally mixed state, namely $\log_{2}d$. Indeed, it
is a less well-known fact that under the most general unitary group,
every density operator can be transformed into the matrix with diagonal
elements uniformly equal to $1/d$ \cite{SM}. $\square$

The assignment of a purely quantum entropic measure $S_{c}(\rho)$
to a state through Eq. (\ref{eq:Sc}) says that the informational
content of a pure state is all coherent, while its (von-Neumann) entropy
is zero. On the opposite side, in a totally mixed state the information
is entirely incoherent. Notice that the sum of $S_{c}(\rho)$ and
$S(\rho)$ is always equal to $\log_{2}d$ for every state, meaning
that every quantum state (at variance with classical ones) produces
a constant unavoidable maximal randomness in the outcomes.

Though the expression in the r.h.s. of (\ref{eq:Sc}) has already
appeared in the literature \cite{Oppenheim2002} as the amount of
thermodynamic work that $\rho$ can extract from a heat bath or the
number of pure state distillable from $\rho$ \cite{Horodeckisetal03},
it was not obtained and interpreted in the present way. In the following
section, we provide another striking interpretation of the same formula. 

\paragraph{Time correlations}

\begin{figure}
\includegraphics[scale=0.3]{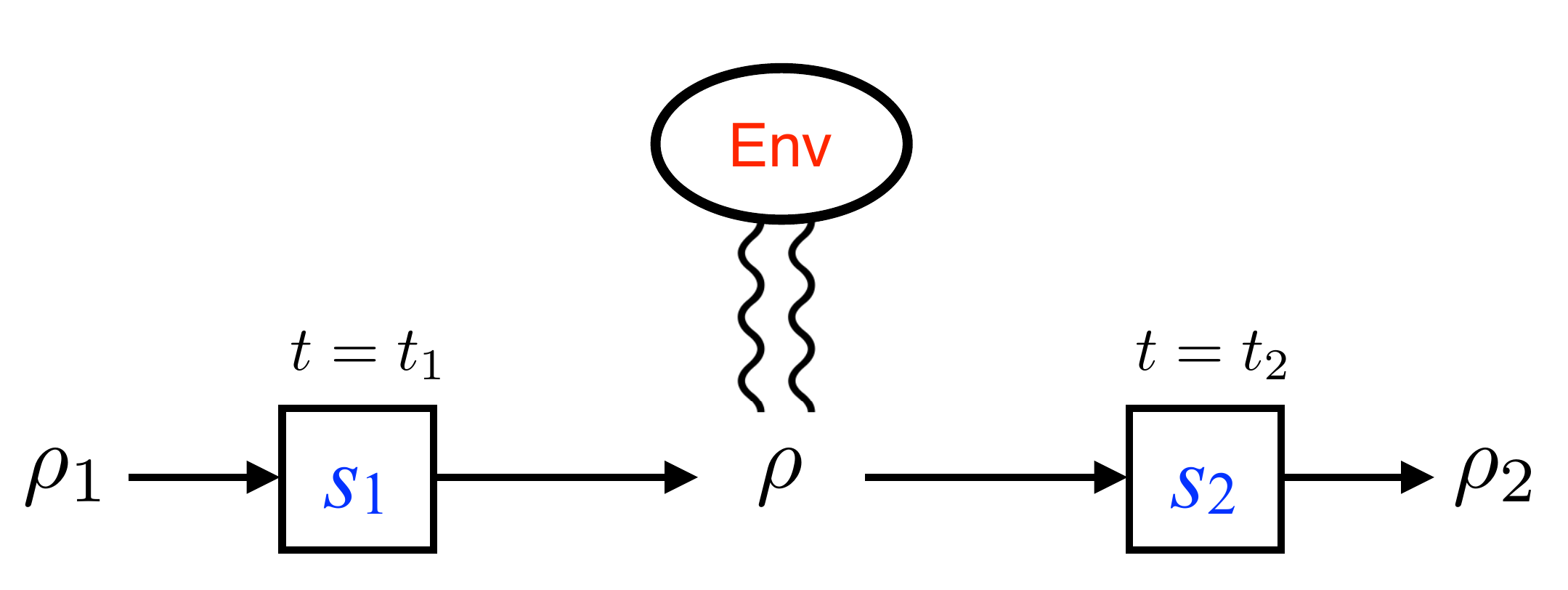}

\caption{Scheme of measurements during time. The state $\rho$ conveys information
between past $s_{1}$ and future $s_{2}$ measurement outputs. Some
information has ``leaked'' into the environment during decoherence,
but the total information is conserved. \label{fig:time scheme} }
\end{figure}

In this section we want to show that the coherent entropy of a given
a state $\rho$ expressed in Eq.(\ref{eq:Sc}) is exactly equal to
the amount of information conveyed between past and future measurements,
due to quantum self-correlations in time. We consider the scheme depicted
in Fig.\ref{fig:time scheme}: the state of interest $\rho$ is prepared
by the measurement of some observable on $\rho_{1}$ at time $t_{1}$
and a subsequent decoherence through interaction with the environment.
At a later time $t_{2}$, the quantum state undergoes another measurement.
The time correlation between the two measurement signals $s_{1}$,
$s_{2}$, with probabilities $p(s_{1})$ and $p(s_{2})$, is estimated
by their mutual information 
\begin{equation}
I_{1:2}=\sum_{s_{1},s_{2}}p(s_{1},s_{2})\log_{2}\left(\frac{p(s_{1},s_{2})}{p(s_{1})p(s_{2})}\right)\label{eq:I12}
\end{equation}
where $p(s_{1},s_{2})$ is the joint probability. The quantum state
can be viewed as a channel, whose capacity is obtained by maximizing
$I_{1:2}$ over all inputs. 

For the sake of clarity, we present here a detailed calculation of
$I_{1:2}$ in the case of one qubit and a depolarizing channel as
a model of decoherence. Assume that initially we have the state $\rho_{1}=\frac{1}{2}(\mathbb{I}_{2}+\mathbf{r}_{1}\cdot\boldsymbol{\sigma})$
represented by the vector $\mathbf{r}_{1}$ inside the Bloch sphere.
At time $t_{1}$, we decide to perform a projective measurement along
the direction described by the unit vector $\hat{\mathbf{n}}_{1}$.
The outcome $s_{1}=\pm1$ will correspond the state $P_{\hat{\mathbf{n}}_{1}}^{s_{1}}=\frac{1}{2}(\mathbb{I}_{2}+s_{1}\hat{\mathbf{n}}_{1}\cdot\boldsymbol{\sigma})$
with probability $\mathrm{Tr}(\rho_{1}P_{\hat{\mathbf{n}}_{1}}^{s_{1}})=\frac{1}{2}(1+s_{1}\mathbf{r}_{1}\cdot\hat{\mathbf{n}}_{1})$.
The subsequent depolarizing channel will simply reduce the length
of the Bloch vector $\hat{\mathbf{n}}_{1}\to\mathbf{n}_{1}$, without
changing its direction. Finally, at time $t_{2}$, we perform a second
projective measurement along $\hat{\mathbf{n}}_{2}$. The outcome
$s_{2}=\pm1$ will be related to the state $P_{\hat{\mathbf{n}}_{2}}^{s_{2}}=\frac{1}{2}(\mathbb{I}_{2}+s_{2}\hat{\mathbf{n}}_{2}\cdot\boldsymbol{\sigma})$
with probability $\mathrm{Tr}(P_{\mathbf{n}_{1}}^{s_{1}}P_{\hat{\mathbf{n}}_{2}}^{s_{2}})=\frac{1}{2}(1+s_{1}s_{2}\mathbf{n}_{1}\cdot\hat{\mathbf{n}}_{2})$.
Hence we have 
\begin{align*}
p(s_{1},s_{2}) & =\frac{1}{2}(1+s_{1}\mathbf{r}_{1}\cdot\hat{\mathbf{n}}_{1})\frac{1}{2}(1+s_{1}s_{2}\mathbf{n}_{1}\cdot\hat{\mathbf{n}}_{2})\\
p(s_{1}) & =\sum_{s_{2}=\pm1}p(s_{1},s_{2})=\frac{1}{2}(1+s_{1}\mathbf{r}_{1}\cdot\hat{\mathbf{n}}_{1})\\
p(s_{2}) & =\sum_{s_{1}=\pm1}p(s_{1},s_{2})=\frac{1}{2}[1+s_{2}(\mathbf{r}_{1}\cdot\hat{\mathbf{n}}_{1})(\mathbf{n}_{1}\cdot\hat{\mathbf{n}}_{2})]
\end{align*}

The mutual information (\ref{eq:I12}) is 
\[
I_{1:2}=H_{2}\left(\frac{1+(\mathbf{r}_{1}\cdot\hat{\mathbf{n}}_{1})(\mathbf{n}_{1}\cdot\hat{\mathbf{n}}_{2})}{2}\right)-H_{2}\left(\frac{1+\mathbf{n}_{1}\cdot\mathbf{\hat{\mathbf{n}}}_{2}}{2}\right),
\]
where $H_{2}(x)=-x\log_{2}x-(1-x)\log_{2}(1-x)$ is the binary entropy.
The maximum of $I_{1:2}$ is achieved for $\mathbf{r}_{1}\cdot\mathbf{\hat{\mathbf{n}}}_{1}=0$
(i.e., the first measurement is orthogonal to the initial state, or
simply the initial state is totally mixed with $\mathbf{r}_{1}=0$)
and $\mathbf{n}_{1}\cdot\mathbf{\hat{\mathbf{n}}}_{2}=|\mathbf{n}_{1}|$
(i.e., the second measurement is collinear to the first one). This
gives the value 
\begin{equation}
I_{1:2}=\log_{2}2-H_{2}\left(\frac{1+|\mathbf{n}_{1}|}{2}\right)=1-S(\rho),\label{eq:I12_max}
\end{equation}
which is the $d=2$ version of $S_{c}(\rho)$ as expressed in Eq.(\ref{eq:Sc}).
Notice that the two possibilities $s_{1}=\pm1$ of intermediate quantum
state $\rho=\frac{1}{2}(\mathbb{I}_{2}+s_{1}\mathbf{n}_{1}\cdot\boldsymbol{\sigma})$
are unitarily equivalent, so they have the same entropy $S(\rho)$.
It is possible to prove the exact match between $S_{c}$ and the maximal
$I_{12}$ also in arbitrary dimension \cite{SM}. 

\paragraph{Conservation of quantum information}

Assuming that the whole universe is in a pure state, then $\rho$
and its environment can be written in Schmidt decomposition and the
entanglement entropy between them is exactly $S(\rho)$ \cite{NielsenChuang_book,Preskill_notes}.
Hence, for every quantum state $\rho$ of dimension $d$ the sum of
the mutual information sent in time \textendash{} quantified by its
coherent entropy $S_{c}(\rho)$ in Eq.(\ref{eq:Sc}) \textendash{}
and the entanglement entropy $S(\rho)$ with the rest of the universe
turns out to be the constant $\log_{2}d$. As a consequence, during
a unitary evolution any loss of coherence is compensated by an equal
increase of entanglement with the environment, and vice versa. This
fact constitutes the basic statement for a \emph{conservation law}
of quantum information. If we interpret $S_{c}(\rho)$ as a measure
of coherence of $\rho$, we obtain that Eq.(\ref{eq:Sc}) is an exact
relation between coherence and entanglement. 

Let us now consider the case of two spatially separated systems $A$
and $B$, described by an overall pure state $\rho_{AB}$. In general
the state of a system is more coherent than the sum of its parts (subadditivity
of entropy), 
\begin{equation}
S_{c}(\rho_{AB})=S_{c}(\rho_{A})+S_{c}(\rho_{B})+I_{A:B}\label{eq:Sc_AB}
\end{equation}
where the excess of coherent entropy amounts to the non-negative quantity
$I_{A:B}=S(\rho_{A})+S(\rho_{B})-S(\rho_{AB})$, which is the (spatial)
mutual information between the two systems $A$ and $B$. Curiously,
the coherent entropy of $\rho_{AB}$ exceeds the sum of the contributions
coming from its parts $A$ and $B$ even when $I_{A:B}$ receives
contribution only from classical correlations. Notice that according
to Eq.(\ref{eq:Sc_AB}) $S_{c}$ obeys the monotonicity property,
at variance with the entropy $S$. Moreover, $S_{c}$ is a convex
function in the space of density matrices. 

Assuming also that $A$ and $B$ are isolated from the rest, so that
unitary operations do not change the value of $S_{c}(\rho_{AB})$,
then we observe that any variation of the ``space-like'' mutual
information $I_{A:B}$ is compensated by an opposite variation of
the ``time-like'' mutual information quantified by the coherent
entropy $S_{c}(\rho_{A})+S_{c}(\rho_{B})$. 

Specifying further to the case of pure $\rho_{AB}$, we fall in the
situation where $B$ is the environment of $A$, and vice versa. Now,
the mutual information $I_{A:B}=2S(\rho_{A})=2S(\rho_{B})$ quantifies
the entanglement between $A$ and $B$, and does not contain any contribution
from classical correlations. The total quantum information consists
of $S_{c}(\rho_{A})$ bits localized in $A$, the same amount in $B$
while the remainder $I_{A:B}$ is encoded in the Hilbert space that
describes both $A$ and $B$. Every unitary process will alter the
balance of these quantities, without changing their sum, which is
equal to the constant $\log d_{A}+\log d_{B}$, i.e. the coherent
entropy of the overall pure state. 

Let us assume that $A$ is localized in a well-defined region in space,
delimited by a closed surface $\Sigma$. The information stored in
$I_{A:B}$ can be assigned to virtual degrees of freedom assigned
to the bonds connecting the real individual subsystems in $A$ and
$B$. Since all these bonds cross the surface $\Sigma$, such information
can be topologically located on it. In other words, during the decoherence
of $A$, also $B$ decoheres, and the consequent lost information
flows from both sides toward the surface: a sort of complementary
of the \emph{holographic principle} known in quantum gravity \cite{Susskind95}.
In this picture, a change in $I_{A:B}$ yields no net ``flow of coherence''
through the surface. At variance with energy, information is a scalar,
so it is relativistically invariant. 

In 1D lattice models, a well-known realization of such mechanism occurs
when we describe matrix-product states (MPS) where the mutual information
between two bipartition $A$ and $B$ of a chain is encoded in the
matrices which describe the bond variables at the border between $A$
and $B$ \cite{Fannes92}.

\paragraph*{Multi-partitions}

After having analyzed the case of two systems, it is interesting to
understand how the quantum information carried by a quantum state
of a given system is distributed when we consider its partition in
several parts \cite{costa2014}. In the case of a tripartition $ABC$,
the overall coherent entropy is given by 

\begin{align}
S_{c}(\rho_{ABC}) & =S_{c}(\rho_{A})+S_{c}(\rho_{B})+S_{c}(\rho_{C})+I_{A:B}+I_{AB:C}\label{eq:S_ABC}
\end{align}
or cyclic permutations of subscripts $A,B,C$. The advantage of having
an expression like Eq.(\ref{eq:S_ABC}), is that it involves only
entropies and mutual informations, which are non negative objects
quantifying amounts of information. The generalization to $n$ partitions
ordered from $1$ to $n$ is 
\[
S_{c}(\rho_{1\cdots n})=\sum_{k=1}^{n}S_{c}(\rho_{k})+I_{1:2}+I_{12:3}+\cdots+I_{1\cdots(n-1):n}
\]
which can be made symmetric with respect to any label ordering. 

\paragraph*{Locally achievable coherence}

The simple result (\ref{eq:Sc}) is obtained when the optimization
problem (\ref{eq:Sc def}) is solved in the space of all the possible
unitary transformations $\mathcal{U}_{\rho}$. However, one may be
interested to restrict the calculation to the family of local transformations
with respect of a given partition. For a bipartite state $\rho_{AB}$
we can define 
\begin{equation}
S_{c}^{loc}(\rho_{AB})=\max_{\sigma\in\mathcal{U}_{\rho_{AB}}^{loc}}\left[\tilde{S}(\sigma)\right]-\min_{\sigma\in\mathcal{U}_{\rho_{AB}}^{loc}}\left[\tilde{S}(\sigma)\right],\label{eq:S_c loc}
\end{equation}
where where $\mathcal{U}_{\rho_{AB}}^{loc}$ is the set of all matrices
which are equivalent to $\rho_{AB}$ under \emph{local} unitaries
$U_{A}\otimes U_{B}$. The result of such an optimization is not guaranteed
to give the same clean expression as in Eq.(\ref{eq:Sc}); instead
we expect a lesser value which must be calculated numerically. It
is appropriate to define the coherence gap $G(\rho_{AB})=S_{c}(\rho_{AB})-S_{c}^{loc}(\rho_{AB})$,
namely the information which cannot be accessed by local operations.
The quantity $G(\rho_{AB})$ accounts for nonlocal correlations between
$A$ and $B$ (not necessarily the entanglement) in a similar fashion
as the quantum discord \cite{discord01}, or the deficit \cite{Horodeckisetal03}.
The remaining local correlations between $A$ and $B$ are quantified
by $L(\rho_{AB})=I_{A:B}-G(\rho_{AB})$. %

\paragraph*{Examples}

In order to familiarize with the concepts discussed in this paper,
we analyze the repartition of information in some quantum states. 
\begin{itemize}
\item The Bell state $|\Psi^{+}\rangle=(|00\rangle+|11\rangle)/\sqrt{2}$
is a pure state with $d=4$, i.e. $S_{c}(\rho_{AB})=2$, meaning 2
bits of information. After partial trace we get $\rho_{A}=\mathrm{Tr}_{B}(|\Psi^{+}\rangle\langle\Psi^{+}|)=\frac{1}{2}(|0\rangle\langle0|+|1\rangle\langle1|)$,
so both the subsystems are totally mixed, with $S_{c}(\rho_{A})=S_{c}(\rho_{B})=0$.
The two bits are stored in the mutual information is $I_{A:B}=2$,
which we can figure out as localized on the whole system $AB$, while
$A$ and $B$ are separately incoherent. Notice that one bit is due
to entanglement entropy, $S(\rho_{A})=1$, while the remaining bit
involves the classical parity correlations \cite{Preskill_notes}.
Remarkably, $G(\rho_{AB})=1$ is the same as the entanglement entropy. 
\item The three-site GHZ state $|\Psi_{GHZ}\rangle=(|000\rangle+|111\rangle)/\sqrt{2}$,
a paradigmatic example where tripartite entanglement is present, while
the pairwise one is zero. The single quantities are summarized in
the following table:\medskip{}
\\
\begin{tabular}{|c|c|c|c|c|c|c|c|}
\hline 
$|\Psi_{GHZ}\rangle$ & $\rho_{A}$ & $\rho_{B}$ & $\rho_{C}$ & $\rho_{AB}$ & $\rho_{AC}$ & $\rho_{BC}$ & $\rho_{ABC}$\tabularnewline
\hline 
\hline 
$S$ & 1 & 1 & 1 & 1 & 1 & 1 & 0\tabularnewline
\hline 
$S_{c}$ & 0 & 0 & 0 & 1 & 1 & 1 & 3\tabularnewline
\hline 
$G$ &  &  &  & 0 & 0 & 0 & 1\tabularnewline
\hline 
$L$ &  &  &  & 1 & 1 & 1 & \tabularnewline
\hline 
$I$ &  &  &  & 1 & 1 & 1 & \tabularnewline
\hline 
$E_{f}$ &  &  &  & 0 & 0 & 0 & \tabularnewline
\hline 
\end{tabular}\medskip{}
\\
where we have also included a row for the entanglement of formation
$E_{f}$, which is exactly computable for pairs of qubits \cite{Wootters98}.
The single subsystems $A$, $B$ and $C$ are all totally incoherent.
Three qubits are stored in the mutual information between pairs, all
made of local correlations. Both the nonlocal indicators $G$ and
$E_{f}$ are vanishing between pairs. Interestingly, $G$ can be computed
also for the tripartite case, resulting in one nonlocally achievable
bit. 
\item The three-site W state $|\Psi_{W}\rangle=(|001\rangle+|010\rangle+|100\rangle)/\sqrt{3}$.
In this case, we have\medskip{}
\\
\begin{tabular}{|c|c|c|c|c|c|c|c|}
\hline 
$|\Psi_{W}\rangle$ & $\rho_{A}$ & $\rho_{B}$ & $\rho_{C}$ & $\rho_{AB}$ & $\rho_{AC}$ & $\rho_{BC}$ & $\rho_{ABC}$\tabularnewline
\hline 
\hline 
$S$ & 0.918 & 0.918 & 0.918 & 0.918 & 0.918 & 0.918 & 0\tabularnewline
\hline 
$S_{c}$ & 0.082 & 0.082 & 0.082 & 1.082 & 1.082 & 1.082 & 3\tabularnewline
\hline 
$G$ &  &  &  & 0.667 & 0.667 & 0.667 & 1.667\tabularnewline
\hline 
$L$ &  &  &  & 0.252 & 0.252 & 0.252 & \tabularnewline
\hline 
$I$ &  &  &  & 0.918 & 0.918 & 0.918 & \tabularnewline
\hline 
$E_{f}$ &  &  &  & 0.550 & 0.550 & 0.550 & \tabularnewline
\hline 
\end{tabular}\medskip{}
\\
Now, some information is carried by single sites, while 0.918 bits
is stored in pairwise correlations: 0.252 local and 0.667 nonlocal.
The presence of nonlocal correlations is confirmed also by 0.550 bits
of entanglement of formation.
\end{itemize}

\paragraph*{Conclusions }

This paper illustrates some arguments which lead to the definition
of an entropic value coming from coherent information in quantum states.
Such a quantity, here called coherent entropy, is indeed physical
as it quantifies the (mutual) information conveyed in time by quantum
states; so it is necessary in order to give a complete description
of their informational content. By means of this quantity and ordinary
mutual information between different systems, it is possible to write
equations of conservation of information in multipartite states, during
unitary processes. Looking at a specific part of an interacting system,
we observe that ``time-like'' information is transformed into ``space-like''
one: the overall information is conserved and the ``flow'' through
a closed surface is governed by a holographic principle. 

As the universe is believed to obey to quantum mechanics where time
evolution is unitary (dissipation and decay processes are not, because
they are only partial descriptions) it is imperative to elevate conservation
of information to a fundamental concept and taking advantage of it,
like it happens with any other conserved quantity. A remarkable consequence
is that no information has been generated or lost since creation of
the universe, but it has only spread out due to expansion and interactions.
The space-time symmetric treatment of mutual information suggests
a possible use in general relativity. For instance, it could help
to shed some light in solving the famous paradox of information loss
in black holes \cite{Susskind95,Braunstein2007}. The change of metric
signature after crossing the event horizon could be responsible of
the transformation of space-like information into time-like, i.e.
a purification of quantum states. This is notoriously connected with
the interpretation of the measurement postulate in quantum mechanics
which invokes a collapse of the wavefunction after extracting some
information about the original state. On the contrary, in the present
framework the consequence of a projective measurement is to \emph{inject}
quantum (coherent) information into a state, as the output quantum
state is pure. 

Finally, it would be interesting to explore other possible consequences
of conservation of coherent information in foundations of quantum
mechanics. We believe that the vision described in the present work
could yield some interesting implications also in field theories and
statistical mechanics. 

\paragraph*{Acknowledgements}

Many thanks to Lorenzo Campos Venuti per very helpful discussions.
This paper is dedicated to the memory of my friend Roberto Ghedini,
who was the most sincere person I ever knew. 

\rule[0.5ex]{1\columnwidth}{1pt}

\section*{Supplemental Material}

\subsection*{Maximal mutual information in time for states in arbitrary dimension }

Referring to the scheme in Fig.\ref{fig:time scheme}, the maximum
of $I_{1:2}$ must be calculated with respect to every measurement
at $t=t_{1}$ and $t=t_{2}$ of the complete observable $A$ and $B$,
respectively. The first measurement $A$ on the initial state $d$-dimensional
$\rho_{\mathrm{in}}$ will generate an output signal $s_{1}=1,\dots,d$,
producing a new state equal to the projector $P_{A}^{s_{1}}=|s_{1}\rangle_{A}\langle s_{1}|_{A}$,
with probability $\mathrm{Tr}(\rho_{\mathrm{in}}P_{A}^{s_{1}})$.
Successively, the system will be subjected to some decoherence, which,
in general, can be modeled in the operator sum formalism as 
\[
\rho=\sum_{m}M_{m}P_{A}^{s_{1}}M_{m}^{\dagger}
\]
with Kraus operators obeying the completeness relation $\sum_{m}M_{m}^{\dagger}M_{m}=\mathbb{I}$.
It is important to consider that in general the intermediate state
$\rho$ depends on the first outcome $s_{1}$. For our scopes, all
these intermediate states should be unitarily equivalent, in order
to correctly quantify the information carried by the given state $\rho.$
Below, we show that such situation is always made possible by a choice
for the $M_{m}$'s. Finally, at time $t_{2}$ the quantum state undergoes
another measurement $B$, generating the output $s_{2}$, with outcoming
state $P_{A}^{s_{2}}$, with probability $\mathrm{Tr}(\rho P_{A}^{s_{2}})$. 

In this process, the joint and marginal probability distributions
are given by
\begin{align*}
p(s_{1},s_{2}) & =\mathrm{Tr}(\rho_{\mathrm{in}}P_{A}^{s_{1}})\mathrm{Tr}(\rho_{s_{1}}P_{B}^{s_{2}})\\
p(s_{1}) & =\mathrm{Tr}(\rho_{\mathrm{in}}P_{A}^{s_{1}})\sum_{s_{2}}\mathrm{Tr}(\rho_{s_{1}}P_{B}^{s_{2}})=\mathrm{Tr}(\rho_{\mathrm{in}}P_{A}^{s_{1}})\\
p(s_{2}) & =\sum_{s_{1}}\mathrm{Tr}(\rho_{\mathrm{in}}P_{A}^{s_{1}})\mathrm{Tr}(\rho_{s_{1}}P_{B}^{s_{2}})
\end{align*}
where we have obviously used the completeness relation $\sum_{s}P_{\alpha}^{s}=\mathbb{I}$,
$\alpha=A,B$. Moreover we have specified the index in the intermediate
state $\rho_{s_{1}}$, remembering that they have the same eigenvectors
and eigenvalues, possibly in different orders. 

So, the mutual information can be written as 
\[
I_{1:2}=\sum_{s_{1},s_{2}}p(s_{1},s_{2})\log_{2}\left(\frac{p(s_{1},s_{2})}{p(s_{1})p(s_{2})}\right)=C_{1}-C_{2}
\]
with
\begin{align*}
C_{1}= & -\sum_{s_{2}}p(s_{2})\log_{2}p(s_{2})\\
C_{2}= & -\sum_{s_{1}}\mathrm{Tr}(\rho_{\mathrm{in}}P_{A}^{s_{1}})\sum_{s_{2}}\mathrm{Tr}(\rho_{s_{1}}P_{B}^{s_{2}})\log_{2}\left(\mathrm{Tr}(\rho_{s_{1}}P_{B}^{s_{2}})\right)
\end{align*}
where $C_{1}$ is the Shannon entropy of the second measurement $H(\{s_{2}\})$,
and $C_{2}$ is the conditional entropy $H(\{s_{2}\}|\{s_{1}\})$.
These two terms can be optimized separately in order to find the maximum
value of $I_{1:2}$ in the space of all the measurement basis $A$
and $B$. Let us observe that $\mathrm{Tr}(\rho_{s_{1}}P_{B}^{s_{2}})$
are just the diagonal elements ($s_{2}=1,\dots,d$) of $\rho_{s_{1}}$
in the $B$ basis. If we indicate with $\tilde{\rho}_{s_{1}}$ the
diagonal part of $\rho_{s_{1}}$ in the $B$ basis, we obtain 
\[
C_{2}=\sum_{s_{1}}\mathrm{Tr}(\rho_{\mathrm{in}}P_{A}^{s_{1}})S(\tilde{\rho}_{s_{1}})
\]
which is clearly minimal when $B$ is the basis of eigenvectors of
all the $\rho_{s_{1}}$'s, giving $C_{2}=S(\rho)$, because all the
$\rho_{s_{1}}$'s have the same eigenvalues. Denoting with $\lambda_{i}$
($i=1,\dots,d$) the eigenvalues of $\rho_{1}$, it is easy to see
that the term $C_{1}$ reaches its theoretical maximum $\log_{2}d$,
with the choice 
\begin{equation}
\begin{cases}
\mathrm{Tr}(\rho_{s_{1}}P_{B}^{s_{2}})=\lambda_{(s_{1}+s_{2}-1)\mathrm{mod}\,d},\\
\mathrm{Tr}(\rho_{\mathrm{in}}P_{A}^{s_{1}})=1/d, & \forall s_{1}=1,\dots,d
\end{cases}\label{eq:choice}
\end{equation}
In other words, the spectra of the density matrices $\rho_{s_{1}}$
are given by all the cyclic permutations of $\{\lambda_{1},\lambda_{2},\dots,\lambda_{d}\}$.
In this way, we get $\sum_{s_{1}}\mathrm{Tr}(\rho_{s_{1}}P_{B}^{s_{2}})=1$,
$\forall s_{2}$. The explicit form of the Kraus operators for obtaining
the first row of Eq. (\ref{eq:choice}) is 

\[
M_{m}=\sum_{s=1}^{d}P_{B}^{m}\Pi_{B}^{s-1}(\Pi_{A}^{s-1})^{\dagger}P_{A}^{s}
\]
with the condition $\mathrm{Tr}(P_{A}^{1}P_{B}^{m})=\lambda_{m}$,
which determines what the basis $A$ should be. The $n$-steps cyclic-permutation
operator of vectors $\{|i\rangle_{A}\}_{i=1}^{d}$ in a given $d$-dimensional
orthogonal basis $A$ is 
\[
\Pi_{A}^{n}=\sum_{k}|k+n\rangle_{A}\langle k|_{A}
\]
Moreover, in the second row of (\ref{eq:choice}) we have required
that $\rho_{\mathrm{in}}$ at $t_{1}$ gives all the possible outcomes
with equal probability. 

In summary, we have obtained that the maximal amount of information
conveyed between past and future measurements is 

\[
I_{1:2}=\log_{2}d-S(\rho)
\]
namely equal to the coherent entropy of $\rho$ expressed in Eq.(\ref{eq:Sc}). 

\subsection*{Every quantum state admit a measurement with completely random outputs}

The statement can be reformulated through the following theorem. 

\textbf{Theorem}. Let $\rho$ be a $d$-dimensional square matrix
with spectrum $\{r_{j},j=1,\dots,d\}$ and related orthonormalized
eigenvectors $|r_{j}\rangle$. Then, it exists a basis of orthonormal
vectors $\{|\phi_{k}\rangle,k=1,\dots,d\}$ where the all the diagonal
elements of $\rho$ are equal. 

\emph{Proof} - We can proceed in a constructive way by writing down
the explicit transformation
\begin{equation}
|\phi_{k}\rangle=\frac{1}{\sqrt{d}}\sum_{j=1}^{d}\exp\left(\frac{2\pi ijk}{d}\right)|r_{j}\rangle.\label{eq:phi_k}
\end{equation}
Let us check that the $|\phi_{k}\rangle$ are orthonormal
\begin{align*}
\langle\phi_{k'}|\phi_{k}\rangle & =\frac{1}{d}\sum_{j,j'=1}^{d}\exp\left(-\frac{2\pi ij'k'}{d}\right)\exp\left(\frac{2\pi ijk}{d}\right)\langle r_{j'}|r_{j}\rangle\\
 & =\frac{1}{d}\sum_{j=0}^{d-1}\exp\left[\frac{2\pi i(k-k')}{d}j\right]=\delta_{kk'}
\end{align*}
which assures the unitarity of the transformation in Eq. (\ref{eq:phi_k}).
The diagonal terms are 
\begin{align*}
\langle\phi_{k}|\rho|\phi_{k}\rangle & =\frac{1}{d}\sum_{j,j'=1}^{d}\exp\left(-\frac{2\pi ij'k}{d}\right)\exp\left(\frac{2\pi ijk}{d}\right)\langle r_{j'}|\rho|r_{j}\rangle\\
 & =\frac{1}{d}\sum_{j=1}^{d}r_{j}=\frac{1}{d}\mathrm{Tr}(\rho)
\end{align*}
for every $k$. 

In particular, if $\rho$ is a density matrix, we have $\mathrm{Tr}(\rho)=1$,
so all the diagonal elements in the basis $|\phi_{k}\rangle$ are
all equal to $1/d$. Notice that the choice (\ref{eq:phi_k}) is not
unique, because we have the freedom to perform the gauge transformation
$|r_{j}\rangle\to\exp(i\theta_{j})|r_{j}\rangle$. 

The result of the present theorem was discussed as a guided exercise
(Problem 3, Sect. 2.2) in the book by R. A. Horn and C. R. Johnson,
\emph{Matrix Analysis}, (Cambridge University Press, 1985) to be solved
through an iterated inverse Jacobi procedure that maximizes the 2-norm
of the off-diagonal part $\sum_{i\neq j}|\rho_{ij}|^{2}$ . 

\cleardoublepage{}
\end{document}